\documentstyle[aps]{revtex}
\newcommand\X{{\vec{x}}}
\newcommand\Y{{\vec{y}}}
\newcommand\T{{{t}}}
\newcommand\dx{d^{4}x}
\newcommand\dy{d^{4}y}
\newcommand\R{{\cal R}}
\newcommand\LL{{|}}
\begin{document}
\twocolumn[\hsize\textwidth\columnwidth\hsize\csname
@twocolumnfalse\endcsname
\title{\Large \bf Relativistic quantum measurement}
\author{Donald Marolf ${}^a$ and Carlo Rovelli ${}^b$\vskip.1cm}
\address{{${}^a$ \it 201 Physics Building Syracuse 
University, Syracuse, NY 13244, USA} \\ 
 {${}^b$ \it Centre de Physique
 Th\'eorique, Luminy, F-13288 Marseille, EU}\\
 { ${}^b$ \it Physics Department, Pittsburgh
 University, PA-15260, USA}}
\date{\today}
\maketitle    
\begin{abstract}
Does the measurement of a quantum system necessarily break Lorentz
invariance?  We present a simple model of a detector that measures the
spacetime localization of a relativistic particle in a Lorentz
invariant manner.  The detector does not select a preferred Lorentz
frame as a Newton-Wigner measurement would do.  The result indicates
that there exists a Lorentz invariant notion of quantum measurement
and sheds light on the issue of the localization of a relativistic
particle.  The framework considered is that of single-particle
mechanics as opposed to field theory.  The result may be taken as
support for the interpretation postulate of the spacetime-states
formulation of single-particle quantum theory.
\end{abstract}  
\vskip1cm]

\section{Introduction}

Does the measurement of a quantum system necessarily break Lorentz
invariance?  Does a state prepared by a quantum measurement
necessarily know about the Lorentz frame in which the measurement was
performed?  Of course, the center of mass of any measurement apparatus
selects a Lorentz frame, and the prepared state may well depend on
this frame.  But can we use a fully Lorentz covariant description of
the system and the apparatus, and formulate a Lorentz covariant
measurement theory, including the projection postulate, such that all
probabilities computed are Lorentz invariant?

The naive Copenhagen-style answer is that a quantum measurement does
break Lorentz invariance: a measurement happens at a certain time $T$,
namely on a specific simultaneity surface.  Therefore it selects a
Lorentz frame.  As a consequence, for instance, the localization of a
quantum relativistic particle is only defined after the choice of a frame.  
One often
discusses the Newton-Wigner position operators \cite{NW}, which are
not covariant.  According to the Newton-Wigner theory, we cannot
simply measure whether or not the particle is at or around a spacetime
point $x$.  We can only measure whether or not the particle is around
$x$ in a certain Lorentz frame.  This is reflected in the fact that
the Newton-Wigner operators in different frames do not commute. 
Accordingly, a (generalized) quantum state prepared by a Newton-Wigner
measurement does not depend only on the spacetime point -- it also
depends on a Lorentz frame at that point.

In the context of field theory, it is of course clear that localized and
covariant measurements can be associated with {\it fields} smeared
over regions of spacetime.  Our concern here is rather with single-particle
mechanics.  In particular, we are interested in the possible implications
for quantum cosmology through its well-known mathematical analogy with
single-particle relativistic mechanics.

Consider then the single-particle setting.  In this framework, can we not 
just measure whether or not the particle is around $x$,
with no reference to a simultaneity surface?  More precisely: isn't it
possible to compute a well defined probability $P_{y,x}$ of detecting the
particle around $y$ if it was previously detected around $x$, such that $P_{y,x}$
would not depend on a preferred Lorentz frame?

In this paper we argue that there is at least one limit in which this
is possible, contrary to what is often assumed (but see
\cite{hartle1}).  Toward this aim, we present a simple model of a
detector for a relativistic particle, and show that the probabilities
of its outcomes are Lorentz invariant.  More precisely, we consider
two detectors.  The first detects the particle in a region $R_x$
around a point $x$ in order to prepare the state for the second
measurement.  Assuming the particle has been detected (and therefore
that its wave function has ``collapsed\footnote{We follow tradition
and use such language, though it is not necessarily our intention to
endorse a Copenhagen interpretation of quantum mechanics.}''), we
calculate the probability that the particle is then detected in a
region $R_y$ around $y$.  We find this probability to be given by a
Lorentz invariant function of $R_x$ and $R_y$.

The two key ingredients for the definition of the detector are as follows.  The
first is the observation \cite{detector,don,hartle2,dh,rr} that any
realistic detector interacts with the system during a time interval
which cannot be null.  Thus, we shall not neglect the finite duration
of the interaction.  Therefore the detection of the particle ``around"
the point $x$ does not mean here in small {\em space} region around
$x$, but rather is a small {\em spacetime} region around $x$.  

The second ingredient is the observation that any physical
interaction, including the one between the system and the measuring
device, must be Lorentz invariant.  Thus we shall choose a Lorentz
invariant interaction Hamiltonian describing the system/apparatus
interaction.  We take these two observations into account and perform
a standard analysis of a measurement using first order perturbation
theory and the standard Copenhagen theory of the wave function
collapse at the time of the measurement.  Our main tool is the
standard trick of exploiting the freedom, pointed out by Von Neumann,
of moving the boundary between the quantum system and the classical
world.  Thus, we describe the apparatus quantum mechanically, and
assume that the Copenhagen measurements happen on the detector. 

The intermediate steps of the calculation are highly noncovariant: the
wave function collapses on a certain simultaneity surface and so on. 
Rather surprisingly the various factors that depend on the Lorentz
frame cancel out at the end.  The result suggests that, at least in
the limit we consider, there exists a Lorentz invariant notion of
quantum measurement and quantum collapse.  One may also choose to take
this as an indication that such an interpretation exists more
generally.

The result also sheds light on the controversial issue of the
localization of a relativistic particle.  The states prepared and
detected by the detector are different from the Newton-Wigner states. 
They were first introduced by Philips \cite{phil}, though without a
measurement interpretation.  The result of this paper therefore shows
that these states do correspond to a rather well defined measurement. 
Unlikely the Newton-Wigner states, the Philips states are defined in a
fully covariant manner.

Finally, a covariant interpretation of quantum theory based on the
so-called spacetime states has been proposed in \cite{rr} (see also
\cite{found}).  This interpretation is based on a covariant
interpretation postulate on the extended configuration space.  In
\cite{rr} it was shown that in the context of non relativistic quantum
mechanics this postulate is equivalent to the standard interpretation. 
The postulate was then assumed to be true, by inference, in more
general contexts.  The problem was raised of whether the postulate
could be reconciled with the predictions of relativistic quantum
particle mechanics.  The result that we obtain here using standard
quantum theory and taking a certain limit is precisely the postulate
of \cite{rr}.  Therefore the result presented here provides some
support to the covariant formulation of quantum theory considered in
\cite{rr}.

\section{A nonrelativistic particle detector}\label{nr}

We begin by describing a related detector in the non-relativistic
context, following \cite{rr} (see also \cite{don} for an earlier
discussion of the same detector).  This serves to set the stage for
our relativistic (and Lorentz invariant) treatment in section IV.

We want to measure the position of the particle at a certain time. 
That is, we want to check whether the particle is present at a certain
space point $\X=0$ at a certain time $\T=0$.  We thus set up a
physical apparatus that interacts with the particle.  This apparatus
will have a pointer that tells us whether or not the particle has been
detected.  We exploit the freedom in choosing the boundary between the
quantum system under observation and the measuring apparatus: we treat
the particle and the detector as the quantum system, and assume that
the Von Neumann measurement is realized when the position of the
pointer is observed.  This trick allows us to better understand which
aspect of the particle state is probed by an apparatus measuring the
localization of the particle.

Consider a pointer which has two possible states.  A state
$|0\rangle$, which corresponds to no detection, and a state
$|1\rangle$, which corresponds to detection.  We represent the state
space of the coupled particle-detector system by the Hilbert space
$H_{PD}=H\otimes C^2$, where $H$ is the Hilbert space of the particle
and $C^2$ is the state space of a two-state system.  We write a state
of the combined system at time $\T$ as (we use the notation
$x=(\X,\T)$)
\begin{equation}
\label{decomp}
\Psi(\T) =  \Psi_0(\T) \otimes |0\rangle +
    \Psi_1(\T) \otimes |1\rangle .  
\end{equation}
At any time after the interaction, one
may describe the two terms in (\ref{decomp})  as ``branches'' of the state
corresponding to detection $(|1\rangle)$ and non-detection $(|0\rangle)$
of the particle.

We write the spacetime wave function of the particle's states
$\Psi_0(\T)$ and $\Psi_1(\T)$ as $\psi_0(\X,\T)=\langle
\X\LL\Psi_0(\T)\rangle$ and $\psi_0(\X,\T)=\langle
\X\LL\Psi_0(\T)\rangle$.  The free Hamiltonian of the particle is
$\vec p^2/2m$, and we take the free Hamiltonian of the detector to be
zero.  Note that here we are in a standard non-relativistic setting so that
the norm of $\Psi_0$ is $\int_{t=t_0} d^dx |\psi_0(x,t_0)|$, where $d$ is
the dimension of a $t=t_0$ slice and, as usual, unitarity guarantees
the norm to be independent of $t_0$.

We need an interaction Hamiltonian $H_{int}$, representing the
interaction that gives rise to the measurement.  $H_{int}$ must have
the following properties.  First, it must cause the transition
$|0\rangle \rightarrow |1\rangle$.  Second, the particle should
interact only at or around the spacetime position ${\X}=0, \T=0$. 
Thus the interaction Hamiltonian must be time dependent, and vanish
for late and early times.  We have to concentrate the interaction
around $\T=0$.  However, we cannot have a perfectly instantaneous
interaction because this would require infinite force to have a finite
effect.  We must
therefore assume that the interaction is non vanishing for a finite
period of time.  Putting these requirements together, and requiring
that the Hamiltonian is self-adjoint, we arrive at an interaction
Hamiltonian of the form
\begin{equation}
 H_{int} = \alpha\ V({\X},{\T})\  \Big(\ |1\rangle \langle 0 | +|0\rangle
 \langle 1 | \ \Big)
\label{inteham}
\end{equation}
where $\alpha\ V({\X},{\T})$ is the potential acting on the particle
in the interaction (with $\alpha$ a coupling constant).  The potential
$V({\X},{\T})$ is concentrated in a small but finite spacetime region
$\cal R$, around ${\X}=0$ and ${\T}=0$.  For simplicity, we take $V$
to be the characteristic function of the region $\R$ (one on $\R$ and
zero elsewhere).  Nothing substantial changes in the discussion if one
uses a different function $V$.

The Schr\"odinger equation for the spacetime wave functions
$\psi_{0}({\X},{\T})$ and $\psi_{1}({\X},{\T})$ reads
\begin{eqnarray}
    \imath\hbar {\partial\psi_{0}\over\partial {\T}}\
    &=&-{\hbar^{2}\over 2m}\ \vec\nabla^2 \psi_{0} +\alpha
    V\ \psi_{1} \\
    \imath\hbar {\partial\psi_{1}\over\partial {\T}}\
    &=&-{\hbar^{2}\over 2m}\vec\nabla^2 \psi_{1} + \alpha
    V\ \psi_{0}.
\end{eqnarray} 

Assume that at some early time ${\T}_{in}<<0$ the particle is in
some arbitrary normalized initial state and the pointer is in the
state $|0\rangle$:
\begin{eqnarray}
\Psi(\T_{in}) & = & \Psi_{0}(\T_{in}) \otimes |0\rangle.
\end{eqnarray} 
What is the state of the system at a later time
${\T}_{f}>>0$?  It is straightforward to integrate the evolution
equations to first order in $\alpha$. One obtains 
\begin{eqnarray}
     \psi_{0}({\X},{\T})\!\! & = &\! \int d{\X}'\
     W({\X},{\T};{\X}',{\T}_{in}) \ \psi_{0}({\X}',{\T}_{in}). 
     \\
     \psi_{1}({\X},{\T}_{f})\!\!  & = & \!  \frac{\alpha}{i\hbar}
     \int \dx'\ W({\X},{\T}_{f};x') \ V(x')\  
      \psi_{0}(x') \\
	& = & \!  \frac{\alpha}{i\hbar}
     \int_{\cal R} \dx'\ W({\X},{\T}_{f};x') \ 
      \psi_{0}(x').
    \label{evolut}
\end{eqnarray}
where $W$ is the propagator 
\begin{eqnarray} 
&& W({\X},\T;{\X}',{\T}') =\nonumber \\
&&\ \ \ \ =\int  \frac{d\vec p}{4\pi^2\hbar^2}\ dE 
\ e^{i/\hbar[\vec p({\X}-{\X}')-E({\T}-{\T}')]}\ \delta(E-\frac{\vec 
p^{2}}{2m})\nonumber \\
&&\ \ \ \ = \int \frac{d\vec p}{4\pi^4\hbar^2} 
\ e^{i/\hbar[\vec p({\X}-{\X}')-\frac{\vec 
p^{2}}{2m}({\T}-{\T}')]} \nonumber \\
&&\ \ \ \ = \left(2\pi m\over i\hbar({\T}-{\T}')\right)^{3\over 2}\
\exp\left\{-{m({\X}-{\X}')^{2}\over 2i\hbar({\T}-{\T}')}\right\}.
\end{eqnarray}
The probability $P_{\R}$ that the pointer is observed in the state
$|1\rangle$ after the interaction is the norm of $\Psi_1(\T_{f})$. 
Using the well known properties of the propagator
\begin{eqnarray}
    \overline{W(x;y)} = W(y;x)
\end{eqnarray}
and
\begin{eqnarray}
   && W({\X},{\T};{\X}',{\T}') = \nonumber \\ 
   &&\ \ \ \ = \int d{\X}'{}'\ 
    W({\X},{\T};{\X}'{}',{\T}'{}') W({\X}'{}',{\T}'{}';{\X}',{\T}'),
\end{eqnarray}
this probability is easily computed
\begin{eqnarray}
P_{\R} &=& \int d\X\ \LL \psi_{1}({\X},{\T}_{f})\LL^2 \nonumber \\
&=&
\frac{\alpha^2}{\hbar^2}
     \int_{\cal R} \dx\ 
     \int_{\cal R} \dx'\ W(x;x') \
     \overline{\psi_{0}(x)}\psi_{0}(x').
\end{eqnarray}
Since we have assumed that $\R$ is small, we can take the lowest order 
terms in the size of $\R$ and assume that $\psi_{0}({\X},{\T})$ is 
constant over $\R$. If $x_{\R}$ is an arbitrary spacetime point in 
$\R$ we have then 
\begin{eqnarray}
P_{\R} &=& 
\frac{\alpha^2}{C^2_{\R}\hbar^2} \ 
     \LL{\psi_{0}(x_{\R})}\LL^2
 \end{eqnarray}
 where
\begin{eqnarray}
\frac{1}{C^2_{\R}}&=& \int_{\cal R} d^4x \int_{\cal R} d^4y\ W(x;y)
\end{eqnarray}
is a normalization factor that plays an important role in what
follows. 

Is the result that we have obtained reasonable?  In order to test it,
let us assume that the region $\R$ has a finite but very small time
extension.  Then the measurement we consider can be identified with a
position measurement at a fixed time, and we must recover the usual
interpretation of the modulus of the wave function as a {\em spatial}
probability density.  If the temporal size $\Delta \T$ of $\R$ is very
small ($m\Delta \T\ll\hbar\Delta V^\frac{2}{3}$) 
compared with its spatial volume $\Delta V$, the normalization
factor ${C^{-2}_{\R}}$ is easy to compute (see \cite{rr}).  It turns
out to be given by
\begin{eqnarray} 
     C_{\cal R}^{-2}  = \Delta V \ \Delta\T^2.
\end{eqnarray}
Therefore the detection probability for this region is 
\begin{eqnarray}
P_{\R} &=& \gamma\ \Delta V\ \LL{\psi_{0}(x_{\R})}\LL^2. 
 \end{eqnarray}
Here 
\begin{eqnarray}
\gamma^2 = \frac{\alpha^2\Delta \T^2}{\hbar^2}
\label{gamma}
\end{eqnarray} 
is a dimensionless parameter that characterizes the efficiency of the
detector.  On the other hand, $ \Delta V\ \LL{\psi_{0}(x_{\R})}\LL^2$
is the probability for the particle to be detected in a small spatial
region of volume $ \Delta V$ at time $\T_{f}$.  Therefore
$\LL\psi_{0}(x_{\R})\LL^2$ is the {\em spatial} probability density and
the result is fully consistent with the standard interpretation of the
wave function.  The factor $\gamma^2$ is interpreted as the intrinsic
efficiency of our detector.  Note that some such parameter is necessarily
present as our perturbative analysis assumes that the interaction is
weak.

After the measurement, we may consider the state of the system collapse to
$\frac{\Psi_{1}}{\LL\LL\Psi_{1}\LL\LL}\otimes|1\rangle$.  Namely after
the measurement, the state of the particle may be described by the
wave function
\begin{eqnarray}
     \psi_{after}(x)\!\!  & = & \!  C_{\R}\ \int_{\cal R} \dy \ W(x;y).
\end{eqnarray}
Notice that the dependences on both the initial wave function and on the 
coupling constant $\alpha$ disappear with the normalization.  We 
denote this state of the particle as $\LL\R\rangle$. That is 
\begin{eqnarray}
     \langle x\LL\R\rangle  & = &  \psi_{after}(x).  
\end{eqnarray}
Explicitly, after the interaction we have
\begin{eqnarray}
     \LL\R\rangle  & = & C_{\R}\int_{\cal R}  \dy\ \LL x\rangle.  
\label{stateR}
\end{eqnarray}
where $\LL x\rangle= \LL\X,\T\rangle$ is the eigenstate of the Heisenberg
position operator $\X(\T)$ with eigenvalue $\vec x$.  This is an
example of a spacetime-smeared state associated to a region, as defined
in \cite{rr}.

Coming back to the state $|\psi_1\rangle \in H$, 
for which $\langle x|\psi_1\rangle = \psi_1(x)$ and which
represents the branch
of the wavefunction in which the particle is detected in ${\cal R}$, 
we see that this state may be written
\begin{equation}
\label{dint}
|\psi_1 \rangle = \gamma  \ \langle {\cal R} | \Psi_0 \rangle  \ | {\cal R} 
\rangle.
\end{equation}
This result is the key to the standard measurement interpretation that the
interaction `measures' some projection associated with the normalized
state $|{\cal R}\rangle$ with some efficiency $\gamma^2$.  It is of course
important that the detector efficiency $\gamma^2$ be independent of the initial
state $|\Psi_0\rangle.$  It immediately follows that the
detection probability $P_{\R}$
can be written as 
\begin{eqnarray}
  P_{\R}  & = & \gamma^2 \ \LL\langle \R \LL \Psi_{0} \rangle \LL^2.
\end{eqnarray}

Summarizing, equation (\ref{dint}) allows us to say that the detector 
we have described prepares
the state $ \LL {\cal R} \rangle$ defined in (\ref{stateR}); the amplitude to
detect an arbitrary $\Psi$ state is $\gamma \langle {\cal R} \LL \Psi
\rangle$, and the efficiency of the detector is $\gamma^2$, given in
(\ref{gamma}).

It is convenient to denote $\langle {\cal R} \LL \Psi \rangle$ as the
amplitude for a particle in the state $\LL \Psi \rangle$ to be
detected in $\R$.  This is the theoretical amplitude of an
hypothetical detector with efficiency 1.  (``Hypothetical" since above
we have used perturbation theory and therefore assumed $\alpha$, and
therefore $\gamma$, to be small.) 

Finally, consider two detectors: the detector 1 in the region
$\R_{1}$, and the detector 2 in the region $\R_{2}$.  We take $\R_{2}$
(entirely) in the past of $\R_{1}$.  Assume that the detector 1 has
detected the particle.  What is then the probability
$P_{\R_{2}\R_{1}}$ that the detector 2 detects the particle?  Applying
the results of the previous section it is immediate to conclude that
the (theoretical: $\gamma=1$) probability is
\begin{eqnarray}
 P_{\R_{2}\R_{1}}  & = & \LL\langle \R_{2} \LL \R_{1} \rangle \LL^2.
\end{eqnarray}
That is 
\begin{eqnarray}
 P_{\R_{2}\R_{1}}  & = & C^2_{R_{1}}C^2_{R_{2}} 
 \left| \int_{R_{1}}\dx\int_{R_{2}}\dy\ W(x;y) \right|^2.
\end{eqnarray}
Equivalently, 
\begin{eqnarray}
   P_{\R_{2}\R_{1}} = \frac{\left|\  W(\R_{2},\R_{1})\ \right|^2} 
 {W(\R_{1},\R_{1})\ W(\R_{2},\R_{2}) }, 
 \label{theorprob}
\end{eqnarray}
where we have defined 
\begin{eqnarray}\label{concl}
   W(\R_{2},\R_{1}) = \int_{R_{1}}\dx\int_{R_{2}}\dy\ \  W(x;y). 
\end{eqnarray}
Of course, this result is in no way Lorentz invariant.

We close this section with two brief comments on these results. 
First, note that this result could also have been achieved by simply
coupling two copies of our detector to the system.  One would then
consider the branch $|\psi_{12}\rangle$ of the state in which both
detectors (the one in ${\cal R}_1$ and the one in ${\cal R}_2$ are
excited).  Each detector has some efficiency $\gamma_1, \gamma_2$
given by the appropriate form of (\ref{gamma}).  {}From (\ref{dint}) it
follows that
\begin{eqnarray}
\langle \psi_{12} | \psi_{12} \rangle &=& \gamma_1^2 \gamma_2^2  \
(\bigl| \langle {\cal R}_1
| \Psi_0 \rangle \bigr|^2) \  \bigl| \langle {\cal R}_1 | {\cal R}_2 \rangle
\bigr|^2 \cr
&=&   \gamma_2^2 \  {\cal P}({\cal R}_1) \  
\bigl|\langle {\cal R}_1 | {\cal R}_2 \rangle \bigr|^2,
\end{eqnarray}
as desired.

Finally, we remind the reader that in order to reach the conclusion
(\ref{concl}) we must ask that ${\cal R}_1$ and ${\cal R}_2$ have a
large separation in time (relative to some scale set by the size of
${\cal R}_2$) so that dispersion does indeed guarantee that the wave
function of the state $|{\cal R}_1\rangle$ is indeed nearly constant
over the region ${\cal R}_2$.

\section{Relativistic dynamics}\label{rp}

The quantum theory of a single relativistic particle is not a
realistic theory since it neglects the physical phenomenon of particle
creation which are described by quantum field theory.  Nevertheless it
is interesting to ask whether there exists a logically consistent
quantum theory, or several, whose classical limit is the dynamics of a
single relativistic particle and which respects the Lorentz invariance
of the classical theory.  Two such quantizations appear natural: one
which contains only positive frequency solutions of the Klein Gordon
equation and one with both frequencies.  For simplicity, we consider
here only the first\footnote{However, our energies will no longer be
entirely positive once we add an interaction with the detector.},
though adding the negative frequency modes should not cause undue
complications.  We start from the classical theory defined by
\begin{eqnarray}
    p^{2} & = & m^{2},
    \label{eq:p2} \\   
    E & > & 0. 
    \label{eq:p0}
\end{eqnarray} 
where $p=(\vec p,E)$ and $p^{2}=-\vec p^{\,2}+E^{2}$.  We use here
$\hbar=c=1$.  Upon quantization, the constraint (\ref{eq:p2}) becomes
the Klein-Gordon equation
\begin{eqnarray}
\Big(\frac{\partial^2}{\partial \T^2}-\vec \nabla^2+m^2\Big) 
\psi(\X,\T)=0 
\end{eqnarray} 
and the positive energy condition (\ref{eq:p0})
becomes the restriction to positive frequencies.  Equivalently, we can
write the relativistic Schr\"odinger equation
\begin{eqnarray}
i\frac{\partial}{\partial
\T}\psi(\X,\T)&=&\sqrt{-\vec\nabla^2+m^2}\ \psi(\X,\T)
\nonumber \\
 &\equiv& H_{0}\ \psi(\X,\T),
\end{eqnarray} 
where the square root is defined by Fourier transform (i.e., by
spectral methods).  The general
solution $\psi(x)$ of these equations is the Fourier transform of
a function supported on the upper mass-$m$ hyperboloid in momentum
space,
\begin{eqnarray}
\psi(x)=\int \frac{d^4p}{4\pi^2}\ \delta(E-\sqrt{\vec p^{\,2}+m^2})\
\tilde\psi(p)\ e^{ipx}.
\label{fourier}
\end{eqnarray} 
Given the wave function $\psi(\X,0)$ on an initial time surface, 
we obtain a solution of the Schr\"odinger equation by
\begin{eqnarray}
\Psi(\T)=e^{-iH_{0}\T}\ \Psi(0) \equiv W_{0}(\T)\ \Psi(0).
\end{eqnarray} 
Explicitly, we have
\begin{eqnarray}
\psi(x)=\int d\X'\ W_{0}(x;\X',0)\ \psi(\X',0), 
\end{eqnarray} 
where the kernel of the evolution operator $W_{0}(\T)$ is the
propagator
\begin{eqnarray}
  W_{0}(x;x')&=&\int \frac{d^4p}{4\pi^2}\ \delta(E-\sqrt{\vec p^2+m^{2}})\ 
  e^{-ip(x-x')} \nonumber\\
  &=& \int \frac{d\vec p }{4\pi^2}\
  e^{i\vec p (\X-\X')-iE(\vec p)(\T-\T')},
\label{prop0}
\end{eqnarray}
with $E(\vec p)=+\sqrt{\vec p^{2}+m^{2}}$.  This propagator is not a
Lorentz invariant object.  For later purposes, we can consider also
the Lorentz invariant propagator
\begin{eqnarray}
  W(x;x')&=&\int \frac{d^{4}p}{4\pi^2}\ \delta(p^{2}-m^{2})\ \theta(E)\
  e^{-ip(x-x')}\nonumber\\
  &=& \int \frac{d\vec p}{4\pi^2}{1\over 2E(\vec p)}\
  e^{iP(\X-\X')-iE(\vec p)(\T-\T')}.
\label{prop}
\end{eqnarray}
Notice that 
\begin{eqnarray}
  W_{0}= 2 H_{0}\ W = (2H_{0})^{\frac{1}{2}}\ W\ (2H_{0})^{\frac{1}{2}}.
  \label{proprel}
\end{eqnarray}
where $W(x;y)$ is the kernel of $W$.

\section{Relativistic particle detector}

Let us now couple a particle detector of the kind considered
in Section \ref{nr} to the relativistic theory described in Section
\ref{rp}.  One may be tempted to simply add the interaction
Hamiltonian
\begin{equation}
 U = \alpha\ V({\X},{\T})\ \Big(\ |1\rangle \langle 0 |
 +|0\rangle \langle 1 | \ \Big)
\label{inteham2}
\end{equation}
to the free positive frequency Hamiltonian $H_{0}$.  But the resulting
theory is not Lorentz invariant (even at the classical level).  
This can easily be seen from the relation
\begin{eqnarray}
    E & = & \sqrt{\vec p\frac{1}{2}^2+m^{2}} + U.  
\end{eqnarray} 
We have (keeping only the liner term in the perturbation)
\begin{eqnarray}
    p^{2} & = & m^{2} + 2EU. 
\end{eqnarray} 

To get a Lorentz invariant theory, we must add a local interaction to
the constraint (\ref{eq:p2}).  That is, we consider instead the interaction between
the particle and the detector defined by
\begin{eqnarray}
    p^{2} & = & m^{2} + U. 
    \label{eq:p2U} \\   
    E & > & 0. 
    \label{eq:p0U}
\end{eqnarray} 
To first order in the coupling we have 
\begin{eqnarray}
  E & = & +\sqrt{\vec p^2+m^{2}} + \frac{U}{2\sqrt{\vec p^2+m^{2}}}.
\end{eqnarray} 
We order the corresponding Schr\"odinger equation symmetrically, 
obtaining the total Hamiltonian 
\begin{eqnarray}
H & = & H_{0} + (2H_{0})^{-\frac{1}{2}}\ U\ (2H_{0})^{-\frac{1}{2}} \equiv 
 H_{0} + H_{int}
\end{eqnarray} 
Therefore, 
\begin{eqnarray}
H_{int} & = &  (2H_{0})^{-\frac{1}{2}}\ U\ (2H_{0})^{-\frac{1}{2}}. 
\label{HintU}
\end{eqnarray} 

Even here the quantum system fails to be manifestly
invariant.  Indeed, we have
\begin{eqnarray}
 -\frac{\partial^2}{\partial \T^2}&&\psi = H^2 \psi  
\cr &&\!\!\! = H_0^2 \psi
+ \frac{1}{2}\left( H_0^{1/2}UH_0^{-1/2} + H_0^{-1/2}UH_0^{1/2}
\right)\!  \psi \cr &&\ \ \  +\  O(U^2).
\end{eqnarray}
However, it turns out that $U$ and $H_0^{-1/2}$ commute in the limit
that we consider.  To see this, note that at the semi-classical level
the commutator is a sum of terms involving derivatives of the characteristic
function $V$ with respect to the spatial 
coordinates $x^i$.  But,  we will act only on states $|\psi_0 \rangle$ 
that are approximately constant over ${\cal R}$, so that 
expectation values involving $\partial_i V$ vanish.  
The vanishing of this commutator
may also be checked by a longer but fully quantum calculation.
As a result, our interaction
is effectively Lorentz invariant.

Let us now consider a setting analogous to that of section \ref{nr},
with the same sort of initial state $|\Psi_0 \rangle$ evolving into a
state with two branches, $|\psi_1\rangle $ and $|\psi_0\rangle$
corresponding to the detection of the particle in ${\cal R}$ and to
the lack of such detection.  If we take $V$ to represent the
Heisenberg operator $V = \int dt V(\vec x, t)$, the branch
corresponding to detection may be written

\begin{equation}
\label{phpsi1}
|\psi_1 \rangle = \frac{\alpha}{\hbar} \frac{1}{\sqrt{2H_0}} V
\frac{1}{\sqrt{2H_0}} |\Psi_0 \rangle.
\end{equation}
Note that the associated wavefunction at a time after the interaction would
contain a factor of $W_0$ (representing time evolution) and would thus
take the same form as in (\ref{evolut}).
In this case, it will turn out that the detection probability is proportional
not to $\Psi_0(x_{\cal R})$, but to $\langle x_{\cal R} | 
\frac{1}{\sqrt{2H_0}} |\Psi_0 \rangle$.  As a result, it is useful to
introduce the state 
\begin{equation}
\label{tilde}
|\tilde  \Psi_0 \rangle = \frac{1}{\sqrt{2H_0}} |\Psi_0 \rangle
\end{equation}
and the associated wave function $\tilde  \Psi_0 (x) = \langle x | \tilde \Psi_0
\rangle.$  

Much as in section \ref{nr}, we now assume that $\tilde
\Psi_0(x)$ is roughly constant
over ${\cal R}$.  In this case, we have $V |\tilde \Psi_0 \rangle
\approx \tilde \Psi_0(x_{\cal R}) \int_{\cal R} d^4 x |x \rangle$ 
so that we may write (\ref{phpsi1}) as
\begin{equation}
|\psi_1 \rangle = \frac{\alpha}{\hbar} \frac{1}{\sqrt{2H_0}} 
\tilde \Psi_0(x_{\cal R}) \int_{\cal R} d^4 x |x \rangle.
\end{equation}
Recall that our goal is to express this in the form (\ref{dint}) of a product
of a state-independent detector efficiency $\gamma$, a normalized state
$|{\cal R}\rangle$, and an inner product
$\langle {\cal R} |\Psi_0 \rangle$ of the initial state with the same
normalized state $|{\cal R} \rangle:$  
\begin{equation}
\label{goodform}
|\psi_1 \rangle = \gamma \ \langle {\cal R} | \Psi_0 \rangle \  |{\cal R} \rangle.
\end{equation}
Introducing the spacetime
volume $V\!ol_4({\cal R})$ of ${\cal R}$ and the normalization factor
\begin{equation}
C_{\cal R}^{-2} = 
W({\cal R, \cal R}) \equiv \int_{\cal R} d^4x \int_{\cal R} d^4y \ 
W(x,y),
\end{equation}
we now make the identifications:
\begin{eqnarray}
\gamma &=& \frac{\alpha}{\hbar C_{\cal R}^2 V\!ol_4({\cal R})}, \cr
| {\cal R} \rangle &=& 
C_{\cal R} \int_{\cal R} d^4x (2H_0)^{-1/2} |x\rangle, \cr
\langle {\cal R} | \Psi_0 \rangle &=& 
C_{\cal R} V\!ol_4({\cal R}) \tilde \Psi_0(x_{\cal R}).
\end{eqnarray}
The last of these identifications is of course not independent, but instead follows
directly from the identification of $|{\cal R} \rangle.$  Note that the 
efficiency (\ref{gamma}) of the detector in section II can also be written
in the above form.

As pointed out in section (\ref{nr}), the form (\ref{goodform})
immediately implies that when
two detectors (associated with regions ${\cal R}_1$ and ${\cal R}_2$) are
considered the probability of detecting the particle in both regions is
\begin{equation}
\gamma_2^2 \  {\cal P}_{{\cal R}_1} \  
|\langle {\cal R}_1 | {\cal R}_2 \rangle |^2,
\end{equation} 
where ${\cal P}_{{\cal R}_1}$ is the probability of detecting the particle
in ${\cal R}_1$.  Thus, idealizing to a perfect ($\gamma =1$) detector of this 
sort, we may say that the probability for a particle 
prepared in $|{\cal R}_1 \rangle$ to arrive in ${\cal R}_2$ is

\begin{equation}
{\cal P}_{{\cal R}_2 {\cal R}_1} = \frac{|W({\cal R}_2,{\cal R}_1)|^2}
{W({\cal R}_1,{\cal R}_1)W({\cal R}_2,{\cal R}_2)}.
\end{equation}
Here 
\begin{eqnarray}
   W(\R_{2},\R_{1}) = \int_{R_{1}}\dx\int_{R_{2}}\dy\ \  W(x;y)
\end{eqnarray}
where $W(x;y)$ is the Lorentz invariant propagator, defined in
(\ref{prop}).  

That is, despite the appearance of $(2H_0)^{-1/2}$ in the definition
of $|{\cal R}\rangle$, the probability amplitude to detect the
particle in $\R_{2}$ if it was detected in $\R_{1}$ is Lorentz
invariant!  In the next section we will come to understand this
factor of $(2H_0)^{-1/2}$ as merely compensating for writing $|{\cal R} \rangle$
in terms of states $|x \rangle$ whose inner product singles out a preferred
Lorentz frame.  This factor will disappear when $|{\cal R} \rangle$ is
written in terms of the truly Lorentz invariant ``Philips states.''

\section{Philips states}

Historically,  two types of (generalized) states have been associated to
spacetime points $x=(\X,\T)$ in relativistic quantum mechanics.  
Recall from equation (\ref{fourier}) that we can write
$\psi(\X,\T)$ as the Fourier transform of a function supported on the
upper mass-$m$ hyperboloid in momentum space 
\begin{eqnarray} 
\psi(x)=\int \frac{d^4p}{4\pi^2}\  \delta(E-\sqrt{\vec p^{\,2}+m^2})\ 
\tilde\psi(p)\ e^{ipx}.
\label{ff} 
\end{eqnarray} 
We can also write the equivalent but more covariant looking expression
\begin{eqnarray} 
\psi(x)=\int d^4p\ \delta(p^2\!-\!m^2)\ \theta(E)\ \tilde\phi(p)\
\sqrt{2p_0}
e^{ipx}.
\label{fourier2}
\end{eqnarray} 
We remind the reader that $\sqrt{2p_0} e^{ipx}$ gives plane
waves with the Lorentz invariant normalization $(2p_0)\delta^{(3)}(\vec p- 
\vec p')$ on the mass shell, corresponding to the Lorentz invariant
measure $d^3p/(2p_0)$.
The relation between (\ref{ff}) and (\ref{fourier2}) being obviously 
\begin{eqnarray} 
\tilde\psi(p) =  \left(\frac{1}{2\sqrt{\vec p^{\,2}+m^2}} \right)^{1/2}
\ \tilde\phi(p).
\end{eqnarray} 

Now, pick a point $y$ in Minkowski space and consider the two
generalized states associated to this point defined, respectively, by
\begin{eqnarray} 
\tilde\psi_{y}(p) =  e^{ipy}.
\end{eqnarray} 
and by 
\begin{eqnarray} 
\tilde\phi_{y}(p) = e^{ipy}.
\end{eqnarray} 
Explicitly, the two states are given by the following two solutions of
the relativistic Schr\"odinger equation
\begin{eqnarray} 
\psi^{\scriptscriptstyle (NW)}_{y}(x)= W_{0}(x,y),
\end{eqnarray} 
and
\begin{eqnarray} 
\psi^{\scriptscriptstyle (PH)}_{y}(x)= W_{1/2}(x,y),
\end{eqnarray} 
where, in operator form,  $W_{1/2} = W_0/(2H_0)^{1/2} = W (2H_0)^{1/2}$.
The Lorentz invariance of the Philips states in now manifest from the 
inner product
\begin{equation}
\langle \psi^{(PH)}_y | \psi^{(PH)}_x \rangle = W(y,x).
\end{equation}

If $y=(\Y,\T)$, the states are given at fixed time $\T$ by 
\begin{eqnarray} 
\psi^{\scriptscriptstyle(NW)}_{y}(\X)=\delta(\X,\Y) 
\end{eqnarray} 
and
\begin{eqnarray} 
\psi^{\scriptscriptstyle(PH)}_{y}(\X)= \int d\vec p\ \left(
\frac{1}{\vec p^{\,2}+m^2}\right)^{1/2}\ e^{i\vec p(\X-\Y)}.
\end{eqnarray} 
Therefore the states $\psi^{\scriptscriptstyle(NW)}_{y}$ form a
(generalized) orthonormal basis
\begin{eqnarray} 
\langle\psi^{\scriptscriptstyle(NW)}_{\Y,\T}\LL\psi^{\scriptscriptstyle(NW)}_{\Y',\T}\rangle= \delta^3(\Y,\Y').
\end{eqnarray} 
while the states $\psi^{\scriptscriptstyle(PH)}_{y}$ do not.  The states
$\psi^{\scriptscriptstyle(NW)}_{y}$ are the well known Newton-Wigner 
states \cite{NW}:
they diagonalize the Newton-Wigner position operator at time $\T$. 
They are non-covariantly defined.  That is, they depend not only on
the spacetime point $y$, but also on the choice of a preferred Lorentz
frame at $y$.  

What about the states $\psi^{\scriptscriptstyle(PH)}_{y}$?  They are
associated to the spacetime point $y$ and are invariantly defined. 
That is, they only depend on the point, not on any choice of reference
frame at the point.  
These states were first considered by Philips \cite{phil}, shortly
after the appearance of the Newton-Wigner paper.  
In spite of the
virtue of being covariantly defined, 
the Philips states have not been very popular. 
The reason is that so far their physical interpretation has not been
clear.  In particular it was not clear what kind of measurement would
produce a Philips state.  The discussion in the previous section
shows that the spacetime detector considered there does indeed prepare
states of this sort.   In particular, 
\begin{eqnarray} 
\LL\R\rangle=\int_{\R}\dx\ |\psi_x^{(PH)}\rangle.
\end{eqnarray} 
An immediate consequence is the property
\begin{eqnarray} 
\langle \psi_y^{(PH)}\LL\psi\rangle= \tilde \psi(y),
\end{eqnarray} 
where $|\tilde \psi\rangle$ is the state introduced in (\ref{tilde}). 
Intuitively, in the limit in which the region $\R$ shrinks to a point
$y$, the states $\LL \R\rangle$ approaches $\LL \psi_y^{(PH)}\rangle$.  
Thus, the detector we have described is a ``detector of Philips 
states". 

Of course, all propagators that we have considered ``propagate faster
than light", as is well known.  They do not vanish at spacelike
separations.  The leakage out of the light cone is small: it is
exponentially damped with the Compton wavelength of the particle.  In
particular, the Philips states associated to the different spatial
points on a given simultaneity surface are not orthogonal to each
other.  This feature of the special relativistic quantum dynamics of a
particle is sometimes regarded as a defect of the theory, which could
compromise its consistency or its classical limit.  We do not think
this is the case.  Simply, the quantum particle has an intrinsic
Compton ``extension" that allows it to excite two spacelike separated
(but close) detectors\footnote{Nevertheless, one wonders if this might
be improved in a single-particle formalism which allows negative frequency
states.}.  In the classical limit, the trajectories stay
inside the light cone.

Of course, this acausal feature makes the theory less attractive than
quantum field theory (in which such effects do not occur).  Note that
this observation implies that the above detectors cannot be
constructed from quantum field theoretic local measuring devices in
any limit.  As a result, they presumably do not correspond to `real'
particle detectors any more than do the Newton-Wigner detectors (which
share this acausal property).  Instead, these detectors exist in a
`relativistic particle' system that is best thought of as a toy model
for quantum cosmology.

\section{Exactly Lorentz invariant detector}

We found the detector above to be effectively Lorentz invariant due to
the fact that $U$ and $H_0$ commute in the limit that we have taken. 
One might ask about a truly Lorentz invariant notion of a spacetime
localized detector.  We shall not discuss this issue in detail, but we
sketch here a possible answer.  

Consider the following manifestly Lorentz invariant algorithm.
By fixing boundary conditions in the past as we did above, one can
compare solutions of the {\it quadratic} constraint $p^2 + m^2 =0$ with
solutions of the perturbed quadratic constraint $p^2 +m^2 + U=0$.  One
simply imposes that the two solutions agree on any Cauchy surface to
the past of the support of $U$.  To the future of the support of $U$,
the perturbed and unperturbed constraints again agree and the
perturbed solution can be written as a sum of two unperturbed
solutions as in (\ref{decomp}).  One would then associate
`probabilities' for detection/non-detection with the norms of these
two unperturbed states.  One needs only a Lorentz-invariant definition
of this norm to complete the discussion.

In general, one cannot restrict consideration to positive frequency
states, as negative frequencies may be introduced by the interaction
$U$.  However, the technique known as `group averaging' (see e.g.
\cite{QORD,ALMMT,GM1,MG}) allows one to define a positive definite
manifestly Lorentz invariant inner product on all solutions of any
constraint of the form $p^2 + m^2 + U = 0$ where $U$ is a localized
disturbance.  In fact, it defines such an inner product in much more
general circumstances as well.  See in particular the recent work of
\cite{Shv1,Shv2,Shv3} for the connection to BRST techniques.  The fact
that it is positive definite is a strong advantage over the
historically more popular Klein-Gordon inner product.

We do not pursue here a detailed treatment of this manifestly
Lorentz-invariant approach, because of the distance from the familiar
von Neumann measurement theory of non-relativistic quantum mechanics. 
Furthermore, due to the assignment of positive norms to negative
frequency states, such a scheme can be physically appropriate only in
the quantum cosmology setting.  In that context, negative frequency
states can be interpreted simply as collapsing universes and not as
particles traveling backwards in time.  Perhaps there is some general
lesson in this last observation, in that one must decide at the outset
whether one wishes to discuss something approximating the relativistic
particles of the real world (which are of course properly described by
excitations of a field theory) or whether one really wishes to discuss
a simplified model of quantum cosmology.  While the two systems seem
rather similar mathematically, the radically different conceptual
status of the associated causal structures on the configuration space
may in the end require radically different foundations for the
corresponding notions of measurement theory and detectors.

\section{Conclusions}

Does a real particle detector detect a Newton-Wigner state or a
Philips state?  Is a real detector better represented by the
interaction that we have described or by a Newton-Wigner operator? 
As noted above, the proper answer is `neither',
as a real particle detector is a local construction in quantum
field theory.  However, taking the relativistic particle as
a toy model for quantum cosmology, one may still ask which detector
is the most useful.  In this context, the Philips detector
has the interesting property of being associated with Lorentz
invariant probabilities.

The interest of the model we have presented, however, is not in the
realism of the model detector considered.  Rather, it is in the fact
that the construction shows that it is possible to think about quantum
measurement in fully a covariant way, at least in a certain limit. 
This result is close in spirit with Hartle's generalized quantum
mechanics \cite{hartle1}.  See also \cite{Gambini,JN,NH}

In particular, the results presented here support the legitimacy of
the particular postulate proposed in \cite{rr,found} for a covariant
spacetime-states formulation of (canonical) quantum
theory\footnote{One can also imagine a fully covariant interpretation
without a class of preferred measurements such as those associated
with spacetime regions above.  In such a framework, our Philips
measurements would stand on an equal footing with the
(frame-dependent) Newton-Wigner measurements.  This is essentially the
formulation used in \cite{QORD}.}.  According to this postulate, the
probability for detecting a system in a small region $\R'$ of the
extended configuration space if it was detected in a small region $\R$
is given by
\begin{eqnarray}
   P_{\R_{2}\R_{1}} = \frac{\left|\  W(\R_{2},\R_{1})\ \right|^2} 
 {W(\R_{1},\R_{1})\ W(\R_{2},\R_{1}) },
\end{eqnarray}
where
\begin{eqnarray}
   W(\R_{2},\R_{1}) = \int_{R_{1}}dx\int_{R_{2}}dy\ \  W(x;y)
\end{eqnarray}
where $dx$ is a measure on the extended configuration space and
$W(x;y)$ is the covariant propagator that defines the quantum theory. 
This postulate is assumed to replace and generalize the usual
interpretation of the wave function, in which measurements happen at
fixed time.  Here we have shown that this postulate is true in a
certain limit of relativistic quantum particle mechanics, provided
that the interaction producing the measurement is described in a
covariant manner.


\begin{thebibliography}{9}

\bibitem{NW} T Newton, E Wigner, Rev Mod Phys 21 (1949) 400. 

\bibitem{hartle1} JB Hartle, {\em Spacetime Quantum Mechanics and the
Quantum Mechanics of Spacetime} in ``Gravitation and Quantizations,
Proceedings of the 1992 Les Houches Summer School", B Julia J
Zinn-Justin eds, Les Houches Summer School Proceedings Vol.  LVII
(North Holland, Amsterdam, 1994).

\bibitem{detector} J Hartle, Phys Rev D44 (1991) 3173.  N Yamada,
S Takagi, Progr Theor Phys 85 (1991) 985.  

\bibitem{don} D Marolf, {\em Models of particle detection in regions
of spacetime}, Phys Rev A 50 (1994) 939.

\bibitem{hartle2} RJ Micanek, JB Hartle, {\em Nearly Instantaneous
Alternatives in Quantum Mechanics\/}, quant-ph/9602023.

\bibitem{dh} JB Hartle, D Marolf {\em Comparing Formulations of
Generalized Quantum Mechanics for Reparametrization Invariant 
Systems}, Phys Rev D56 (1997) 6247-6257.

\bibitem{rr} M Reisenberger, C Rovelli, {\em Spacetime states and
covariant quantum theory}, gr-qc/0111016

\bibitem{phil} TO Philips, Phys Rev 136 (1964) B893.

\bibitem{found} C Rovelli, {\em A note on the foundation of
relativistic mechanics}, gr-qc/0110075

\bibitem{QORD} D Marolf
{\em Quantum observables and recollapsing dynamics},
Class Quant Grav 12 (1995) 1199, gr-qc/9404053.

\bibitem{ALMMT}
A Ashtekar, J Lewandowski, D Marolf, J Mour\~ao, and T Thiemann, {\em
Quantization of diffeomorphism invariant theories of connections with
local degrees of freedom}, J.~Math.  Phys.  {\bf 36} (1995) 6456,
gr-qc/9504018.
 
\bibitem{GM1} D Giulini, D Marolf, {\em On the Generality of
Refined Algebraic Quantization}, Class Quant Grav 16 (1999) 2479,
gr-qc/9812024.

\bibitem{MG} D Marolf, {\em 
Group Averaging and Refined Algebraic Quantization: Where are we now?}, 
In the proceedings of the 9th Marcel-Grossman Conference, Rome 2000;
gr-qc/0011112.

\bibitem{Shv1} O Shvedov, {\em BRST-BFV, Dirac and Projection
Ope\-rator Quantizations: Correspondence of States}, hep-th/0106250.

\bibitem{Shv2} O Shvedov, {\em Refined Algebraic Quantization of
Constrained Systems with Structure Functions}, hep - th/ 0107064.

\bibitem{Shv3} O Shvedov,
{\em On Correspondence of BRST-BFV, Dirac and Refined 
Algebraic Quantizations of Constrained Systems}, hep-th/0111270.

\bibitem{Gambini} R Gambini, RA Porto {\em Relational time in
generally covariant quantum systems: four models}, Phys Rev D63
(2001) 105014; {\em Relational Reality in Relativistic Quantum
Mechanics}, quant-ph/0105146.

\bibitem{JN} JJ Halliwell, J Thorwart, {\em Decoherent histories
analysis of the relativistic particle}, Phys Rev D64 (2001) 124018. 
JJ Halliwell, {\em Life in an Energy Eigenstate: Decoherent Histories
Analysis of a Model Timeless Universe}, gr-qc/0201070.


\bibitem{NH} H Nikoli\'c
{\em A general-covariant concept of particles in curved 
background}, Phys Lett B527 (2002) 119-124. 

\end{thebibliography}
\end{document}